\let\ifarxiv=\iffalse    % JOURNAL VERSION
\pdfoutput=1

\ifarxiv

\documentclass[12pt,a4paper]{article}
\usepackage[a4paper,text={450pt,650pt},centering]{geometry}

\fi

\ifarxiv\else

\documentclass[11pt,a4paper]{article}
\usepackage{mathptmx}
\usepackage[a4paper,text={130mm,198mm}]{geometry}

\fi

\usepackage{amsmath,amssymb}
\usepackage[latin2,latin9]{inputenc}
\usepackage{xcolor}
\usepackage{pdfcolmk}
\usepackage{graphicx}
\usepackage{esint}
\providecommand{\tabularnewline}{\\}
\providecolor{lyxadded}{rgb}{0,0,1}
\providecolor{lyxdeleted}{rgb}{1,0,0}

\usepackage[bookmarks=true,hyperfigures=true]{hyperref}
\usepackage{graphicx}
\usepackage[nosort]{cite}
\let\oldbfseries=\bfseries
\let\oldmdseries=\mdseries
\let\oldnormalfont=\normalfont
\renewcommand{\bfseries}{\oldbfseries\boldmath}
\renewcommand{\mdseries}{\oldmdseries\unboldmath}
\renewcommand{\normalfont}{\oldnormalfont\unboldmath}

\allowdisplaybreaks[3]

\numberwithin{equation}{section}

\usepackage[font=small,labelfont=bf,width=0.85\textwidth]{caption}

\providecommand{\hypersetup}[1]{}

\hypersetup{plainpages=false}
\hypersetup{pdfpagemode=UseNone}
\hypersetup{bookmarksnumbered=true}
\hypersetup{pdfstartview=FitH}
\hypersetup{colorlinks=false}
\hypersetup{citebordercolor={.5 1 .5}}
\hypersetup{urlbordercolor={.5 1 1}}
\hypersetup{linkbordercolor={1 .7 .7}}
\providecommand{\arxivref}[2]{\href{http://arxiv.org/abs/#1}{#2}}
\providecommand{\doiref}[2]{\href{http://dx.doi.org/#1}{#2}}

\providecommand{\href}[2]{#2}
\providecommand{\arxivlink}[1]{\href{http://arxiv.org/abs/#1}{arxiv:#1}}

\begin{document}

%%%%%%%%%%%%%%%%%%%%%%%%%%%%%%%%%%%%%%%%%%%%%%%%%%%%%%%%%%%%%%%%%%%%%%%%%%%%%%%%
%%%%%%%%%%%%%%%%%%%%%%%%%%%%%%%%%%%%%%%%%%%%%%%%%%%%%%%%%%%%%%%%%%%%%%%%%%%%%%%%
% TITLE PAGE

\thispagestyle{empty}
\phantomsection
\addcontentsline{toc}{section}{Title}

\begin{flushright}\footnotesize%
\texttt{\arxivlink{1012.3995}}\\
overview article: \texttt{\arxivlink{1012.3982}}%
\vspace{1em}%
\end{flushright}

\begingroup\parindent0pt
\begingroup\bfseries\ifarxiv\Large\else\LARGE\fi
\hypersetup{pdftitle={Review of AdS/CFT Integrability, Chapter III.6: Thermodynamic Bethe Ansatz}}%
Review of AdS/CFT Integrability, Chapter III.6:\\
Thermodynamic Bethe Ansatz
\par\endgroup
\vspace{1.5em}
\begingroup\ifarxiv\scshape\else\large\fi%
\hypersetup{pdfauthor={Zolt\'an Bajnok}}%
Zolt\'an Bajnok
\par\endgroup
\vspace{1em}
\begingroup\itshape
\emph{Theoretical Physics Research Group of the }\\
\emph{Hungarian Academy of Sciences, }\\
\emph{H-1117 P\'azm\'any s. 1/A, Budapest, Hungary}
\par\endgroup
\vspace{1em}
\begingroup\ttfamily
bajnok@elte.hu
\par\endgroup
\vspace{1.0em}
\endgroup

\begin{center}
\includegraphics[width=5cm]{TitleIII6.mps}%figure for your chapter
\vspace{1.0em}
\end{center}

\paragraph{Abstract:}
The aim of the chapter is to introduce in a pedagogical manner the
concept of Thermodynamic Bethe Ansatz designed to calculate the energy
levels of finite volume integrable systems and to review how it is
applied in the planar AdS/CFT setting. 

\ifarxiv\else
% http://www.ams.org/msc
\fi
\hypersetup{pdfsubject={MSC (2010): 81T30, 81T40, 81U15, 81U20 }}%

\ifarxiv\else
\fi
\hypersetup{pdfkeywords={Thermodynamic Bethe Ansatz, planar AdS/CFT}}%

\newpage

%%%%%%%%%%%%%%%%%%%%%%%%%%%%%%%%%%%%%%%%%%%%%%%%%%%%%%%%%%%%%%%%%%%%%%%%%%%%%%%%
%%%%%%%%%%%%%%%%%%%%%%%%%%%%%%%%%%%%%%%%%%%%%%%%%%%%%%%%%%%%%%%%%%%%%%%%%%%%%%%%
% BODY

%%%%%%%%%%%%%%%%%%%%%%%%%%%%%%%%%%%%%%%%%%%%%%%%%%%%%%%%%%%%%%%%%%%%%%%%%%%%%%%%
\section{Introduction}

Thermodynamic Bethe Ansatz (TBA) is a method to calculate exactly
the groundstate energy of an integrable quantum field theory in finite
volume using its infinite volume scattering data \cite{Zamolodchikov:1989cf}%
\footnote{The method has its origin in the work of Yang and Yang applied for
spin chains and for the Bose gas with $\delta$ interaction \cite{Yang:1968rm}.%
}. The equations can be extended to excited states as well by analytical
continuation \cite{Dorey:1996re,Dorey:1997rb}.

The idea of the TBA is to exploit that the Euclidean partition function
is dominated for large imaginary times by the groundstate energy.
Calculating the partition function in the doubly Wick rotated (mirror)
theory the imaginary time becomes the physical size which is taken
to be large. Since the large volume spectrum is under control, the
partition function can be evaluated in the saddle point approximation
which results in nonlinear integral equation for pseudo energies leading
to an exact description of the ground state energy. Excited states
on the complex (volume/coupling) plane are connected to the groundstate
which enables one to derive nonlinear integral equations for excited
states as well. 

We start in Section 2 with a toy model containing one single particle
with AdS dispersion relation and with scattering matrix which is not
a function of the differences of the momenta. Although this is a fictitious
system it helps to introduce conceptual notions and steps needed to
explain the TBA which is, in analogy, used in Section 3 to present
the results for planar AdS/CFT. Finally, we give a guide to the literature
in Section 4 and list some open problems.

\section{The concept of TBA: a toy model}

The application of the TBA method to solve completely the finite volume
spectral problem is standard by now and follows the following steps.
First the scattering theory has to be solved in infinite volume by
determining the scattering matrix from its generic properties such
as symmetry, unitarity, crossing relation. The poles of the scattering
matrix lying in the physical strip are related to bound-states. These
bound-states have to be mapped and their scattering matrices have
to be determined from the constituents' scattering matrices. Then
in the second step these scattering matrices can be used to describe
the spectrum for large volume, which amounts to restrict the allowed
particles' momenta via phase shifts and periodicity, and use the dispersion
relation to express the energy in terms of the quantized momenta.
This method sums up all power like corrections in the inverse of the
volume and provides an asymptotical spectrum. The very same asymptotic
description of the mirror theory is also needed as it can be used
to calculate the exponentially small finite energy corrections from
the partition function. Evaluating the Euclidean partition function
for large imaginary times (large mirror volumes) in the saddle point
approximation provides integral equations describing the ground state
energy exactly. Finally, these equations can be extended for excited
states by analytical continuation. Now let us see how these steps
are elaborated in the simplest setting.

\subsubsection*{Infinite volume characteristics of the model}

We consider a toy model with one single particle type only. The dispersion
relation is supposed to be the same as in the AdS/CFT correspondence
\footnote{The string tension is related to the 't Hooft coupling as $2\pi g=\sqrt{\lambda}$
}:\[
E(p)=\sqrt{1+4g^{2}\sin^{2}\frac{p}{2}}\]
The sine function indicates lattice behavior and restricts the momentum
as \break $p\in[-\pi,\pi]$. The square root, however, has a relativistic
origin. The theory is supposed to be integrable, thus multiparticle
scattering matrices factorize into two particle scatterings. As relativistic
invariance is not supposed the two particle S-matrix can depend separately
on the two momenta $S(p_{1},p_{2})$ and satisfies unitarity $S(p_{1},p_{2})S(p_{2},p_{1})=1$
and crossing symmetry, which helps to fix it completely. We will not
need its explicit form, but will suppose that in the $p_{1}=p_{2}=p$
particular case $S(p,p)=-1$.

\subsubsection*{Infinite volume characteristics of the mirror model}

The Euclidean version of the model is defined by analytically continuing
in the time variable $t=iy$ and considering space $x$ and imaginary
time $y$ on an equal footing. The Euclidean theory so obtained can
be considered as an analytical continuation of another theory, in
which $x$ serves as the analytically continued time $x=i\tau$ and
$y$ is the space coordinate. The theory defined in terms of $y,\tau$
is called the mirror theory and its dispersion relation can be obtained
by the same analytical continuation $E=i\tilde{p}$ and $p=i\tilde{E}$
which results in\[
\tilde{E}(\tilde{p})=2\mbox{arcsinh}\left(\frac{1}{2g}\sqrt{\tilde{p}^{2}+1}\right)\]
Contrary to the original theory the mirror model is not of the lattice
type as its momentum can take any real value $\tilde{p}\in\mathbb{R}.$
As the scattering matrix is related via the reduction formula to the
Euclidean correlator the mirror S-matrix is simply the analytical
continuation of the original scattering matrix: $S(\tilde{p}_{1},\tilde{p}_{2})$.

\subsubsection*{Very large volume solution: asymptotic Bethe Ansatz for the model}

Let us put $N$ particles in a large volume $L$ subject to periodic
boundary condition. Integrability ensures that the particle number
is conserved and the particles' momenta are not changed in the consecutive
scatterings. The leading effect of the finite volume is the momentum
quantization constraint: 
\begin{equation}
1=e^{ip_{j}L}\prod_{k:k\neq j}^{N}S(p_{j},p_{k})\label{eq:ABA}
\end{equation}
which is called the Bethe Yang equation or asymptotic Bethe Ansatz
(ABA) and follows from the periodicity of the multiparticle wave function.
Due to the sine function in the dispersion relation and the periodicity
of $p$ consistency of (\ref{eq:ABA}) requires $L$ to take integer
values only. 

Bound-states of the theory are manifested in the ABA as complex string-like
solutions. Indeed, if the scattering matrix has a pole for $\Im m(p)>0$,
then complex $p$ solutions are also allowed in (\ref{eq:ABA}). If
we take $L$ very large with $p_{1}\approx\frac{p}{2}+iq$ then the
rhs. of (\ref{eq:ABA}) would go to $\infty$ which should be compensated
by another complex momentum, $p_{2}\approx\frac{p}{2}-iq$ say, such
that $S(p_{1},p_{2})$ exhibits a pole. The two particles with momenta
$p_{1}$ and $p_{2}$ form a bound-state with momentum $p=p_{1}+p_{2}$,
energy $E_{2}(p)=E(p_{1})+E(p_{2})$ and scattering matrix $S_{21}(p,p_{j})=
S(\frac{p}{2}+iq,p_{j})S(\frac{p}{2}-iq,p_{j})$.
In general complex solutions built up from more particles are also
allowed and they usually form a string-like pattern. Their dispersion
relation and scattering matrices can be calculated by extending the
method above, which is called the S-matrix bootstrap.

\subsubsection*{Very large volume solution: ABA for the mirror model}

In the mirror model the considerations go along the same line as in
the original theory. If we denote the mirror volume by $R$ the ABA
reads as
\begin{equation}
1=e^{i\tilde{p}_{j}R}\prod_{k:k\neq j}S(\tilde{p}_{j},\tilde{p}_{k})
\label{eq:MirrorABA}
\end{equation}
Since $S(\tilde{p}_{1},\tilde{p}_{2})$ lives in a different analytical
domain than $S(p_{1},p_{2})$ its pole structure can be also different.
If it exhibits poles also at the proper location the mirror theory
has also bound-states. Once bound-states exist we can calculate their
dispersion relation and scattering matrices from the bootstrap method.
Suppose that the bound-states can be labeled with some charge $Q$,
they have energy $\tilde{E}_{Q}(\tilde{p})$ and their scattering
matrix is $S_{Q_{j}Q_{k}}(\tilde{p}_{j},\tilde{p}_{k})$. The generic
ABA valid for all the excitations (also for bound-states) will be
\begin{equation}
1=e^{i\tilde{p}_{j}R}\prod_{k:k\neq j}S_{Q_{j}Q_{k}}(\tilde{p}_{j},\tilde{p}_{k})
\label{eq:MirrorQABA}
\end{equation}
Once these equations are solved the energy of the multiparticle state
is\[
\tilde{E}=\sum_{j=1}^{N}\tilde{E}_{Q_{j}}(\tilde{p}_{j})\]
which describes the spectrum asymptotically for large volumes $R$.

\subsubsection*{Groundstate TBA equation from the partition function}

Let us come back to the original model and see how the exact groundstate
energy can be determined in a finite volume $L$ from the Euclidean
partition function. We exploit the fact that the imaginary time evolution
for large times, $R$, is dominated by the lowest energy state
\[
\lim_{R\to\infty}Z(L,R)=\lim_{R\to\infty}
\mbox{Tr}(e^{-RH(L)})=\lim_{R\to\infty}e^{-RE_{0}(L)}+\dots
\]
where the ellipsis represents terms exponentially suppressed in $R.$
The same partition function can be determined alternatively, using
the time evolution of the mirror theory which is generated by the
mirror Hamiltonian $\tilde{H}$:
\[
Z(L,R)=\tilde{Z}(R,L)=\mbox{Tr}(e^{-L\tilde{H}(R)})=\sum_{n}e^{-L\tilde{E}_{n}(R)}
\]

\begin{table}
\begin{centering}
\begin{tabular}{|c|c|}
\hline 
original model & mirror model\tabularnewline
\hline
\hline 
$(t,x)\equiv(y=it,x)$ & $(y,x=i\tau)\equiv(\tau,y)$\tabularnewline
\hline 
\includegraphics[width=3cm]{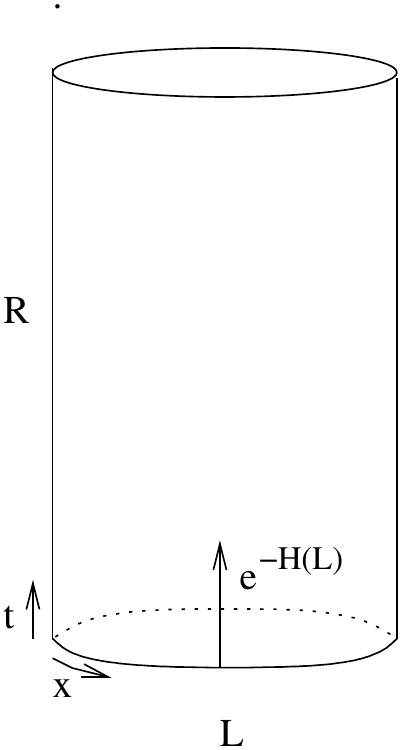} & \includegraphics[height=3cm]{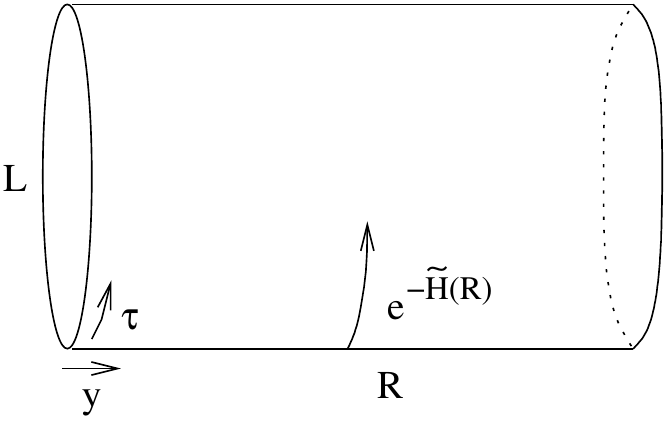}\tabularnewline
\hline 
$(E,P)\equiv(P_{y}=iE,P_{x}=P)$ & $(P_{y}=\tilde{P},P_{x}=i\tilde{E})\equiv(\tilde{E},\tilde{P})$
\tabularnewline
\hline
\end{tabular}
\par\end{centering}

\caption{The relation between the original and the mirror model. }
\label{Flo:orig/mirror}
\end{table}

The relation between the original model and the mirror model is summarized
in Table \ref{Flo:orig/mirror}. In switching to the mirror model we ensure that the volume
goes to infinity (and not the imaginary time) where the spectrum is controlled by the ABA
(\ref{eq:MirrorQABA}). 

In the large $R$ limit the sum in the partition function is dominated
by finite density particle states. Introducing the density of the
particles (and bound-states) in momentum space 
($\rho_{Q}(\tilde{p})=\frac{\Delta n_{Q}}{R\Delta\tilde{p}}$)
the energy can be expressed as
\[
\tilde{E}[\rho]=R\sum_{Q}\int d\tilde{p}\,\rho_{Q}(\tilde{p})\,
\tilde{E}_{Q}(\tilde{p})=R\sum_{Q}\int du\,\rho_{Q}(u)\,\tilde{E}_{Q}(u)
\]
where for later convenience we reparametrized the momentum as $\tilde{p}(u)$,
momentum integrations go from $-\infty$ to $\infty$. The quantization
condition comes from taking the logarithm of the mirror ABA
\begin{equation}
\tilde{p}_{j}(u_{j})+\sum_{Q'}\int du'(-i\log S_{Q_{j}Q'}(u_{j},u'))
\rho_{Q'}(u')=2\pi\frac{n_{j}}{R}\label{eq:rhoABA}
\end{equation}
where $n_{j}$ labels the quantized momentum $\tilde{p}_{j}$ whose
charge is $Q_{j}$. For a generic multiparticle state there are momenta
$\tilde{p}_{k}$ which satisfy the same equation but which are not
excited, not present in the system. They are called holes and their
densities in the large volume limit is described by $\bar{\rho}_{Q}.$
Clearly the densities of particles and holes are not independent they
are connected by the thermodynamical limit of eq. (\ref{eq:rhoABA})
as
\begin{equation}
\partial_{u}\tilde{p}-2\pi(\rho_{Q}+\bar{\rho}_{Q})=
-\sum_{Q'}\int du'K_{QQ^{'}}(u,u')\rho_{Q'}(u')=:-K_{QQ^{'}}\star\rho_{Q'}\label{eq:rhobarrhoABA}
\end{equation}
where the kernel is defined as\[
K_{QQ^{'}}(u,u')=-i\partial_{u}\log S_{QQ'}(u,u')\]
The particle density itself does not characterize properly the states
we sum over in the partition function. Indeed in a given interval
$(u,u+du)$ the occupied $R\rho_{Q}(u)du$ particles can be distributed
${R(\rho_{Q}(u)+\bar{\rho}_{Q}(u))du \choose R\rho_{Q}(u)du}$ different
ways leading to an entropy factor in the sum. Since in the large particle
number limit the factorials can be approximated with the Stirling
formula the partition function will take the form\[
Z(L,R)=\sum_{n}e^{-L\tilde{E}_{n}(R)}=\sum_{Q}\int d[\rho_{Q}]e^{-L\tilde{E}[\rho_{Q}]
+S[\rho_{Q},\bar{\rho}_{Q}]}\]
where the entropy factor is\[
S[\rho_{Q},\bar{\rho}_{Q}]=R\int du\left[(\rho_{Q}+\bar{\rho}_{Q})
\log(\rho_{Q}+\bar{\rho}_{Q})-\rho_{Q}\log\rho_{Q}-\bar{\rho}_{Q}\log\bar{\rho}_{Q}\right]\]
One can slightly generalize the partition function by adding a chemical
potential term to the energy $-L\tilde{E}_{Q}[\rho_{Q}]\to\mu_{Q}[\rho_{Q}]-L\tilde{E}_{Q}[\rho_{Q}]$
where $\mu_{Q}[\rho_{Q}]=R\mu_{Q}\sum_{Q}\int du\,\rho_{Q}(u)$. For
fermions we take $\mu_{Q}=i\pi$ , while for bosons $\mu_{Q}=0$.
This extended partition function can be evaluated in the saddle point
approximation. Taking into account the relation between $\delta\rho_{Q}$
and $\delta\bar{\rho}_{Q}$ originating from the variation of (\ref{eq:rhobarrhoABA})
we obtain the minimizing equation in the so called pseudo energy $\epsilon_{Q}=\log\frac{\bar{\rho}_{Q}}{\rho_{Q}}$
as
\begin{eqnarray*}
\epsilon_{Q}(u)-L\tilde{E}_{Q}(u)+\mu_{Q} & = & -\sum_{Q'}\int
\frac{du'}{2\pi}K_{Q'Q}(u',u)\log(1+e^{-\epsilon_{Q'}(u')})\\
 & =: & -(\log(1+e^{-\epsilon_{Q'}})\star K_{Q'Q})(u)\end{eqnarray*}
Once the pseudo energies are determined the ground state energy in
volume $L$ can be obtained as\begin{equation}
E_{0}(L)=-\sum_{Q}\int\frac{du}{2\pi}(\partial_{u}\tilde{p})\,\log(1+e^{-\epsilon_{Q}(u)})
\label{eq:TBAenergy}
\end{equation}
The nonlinear integral equation which determines the pseudo energies
is called the thermodynamic Bethe Ansatz (TBA) equation. Although
it is not possible to solve it in general it provides an implicit
exact description of the groundstate energy. This implicit solution
is a starting point of a systematic large and small volume expansion
and can be used to derive either functional relations for the pseudo
energies or TBA equations for excited states by analytical continuation.

\subsubsection*{Excited states by analytical continuation}

Here we start with bosonic theories without bound-states and suppose
that by analytically continuing in some parameter (say in the volume)
we can reach all excited states. The way how excited states appear
can be understood by analyzing the energy expression (\ref{eq:TBAenergy})
integrated by parts\[
E=\int\frac{du}{2\pi}\tilde{p}(u)\partial_{u}\log(1+e^{-\epsilon(u)})\]
Let us suppose that in the analytical continuation singularities of
type $1+e^{-\epsilon(u_{i})}=0$ appear. When we deform the contour
their residue contributions give rise to\[
E=\sum_{i}E(u_{i})-\int\frac{du}{2\pi}\partial_{u}\tilde{p}(u)\log(1+e^{-\epsilon(u)})\]
where we took into account the relation between the energy and the
mirror momentum $E(u_{j})=i\tilde{p}(u_{j})$. Taking the same analytical
continuation in the equation for the pseudo energy we obtain\[
\epsilon(u)=L\tilde{E}(u)+\sum_{i}\log S(u_{i},u)-\int\frac{dw}{2\pi}K(w,u)\log(1+e^{-\epsilon(w)})\]
Solving these equations iteratively for large $L$ we can recognize
that the $1+e^{-\epsilon(u_{i})}=0$ equations, which determine the
positions of the singularities, coincide at leading order with the
ABA equations (\ref{eq:ABA}). The subleading order calculation provides
a universal formula for the leading finite size correction of multiparticle
energy levels \cite{Bajnok:2009vm}. Alternatively for doing the analytical
continuation one can think of the final result as choosing a different
integration contour which surrounds the $1+e^{-\epsilon(u_{i})}=0$
singularities, and when we take the integration contour back to the
real axis we pick up the above residue contributions. 

Finally we note that if we have more species (labeled by $Q$) with
diagonal scatterings (like in the previous subsection) then a singularity
in $1+e^{-\epsilon_{Q_{i}}(u_{i})}=0$ results in the equations\[
\epsilon_{Q}(u)=L\tilde{E}_{Q}(u)+\sum_{i}\log S_{Q_{i}Q}(u_{i},u)
-(\log(1+e^{-\epsilon_{Q'}})\star K_{Q'Q})(u)\]
whose solutions $\epsilon_{Q}(u)$ and $\{u_{i}\}$ have to be plugged
into the energy formula\[
E=\sum_{i}E_{Q_{i}}(u_{i})-\sum_{Q}\int\frac{du}{2\pi}\partial_{u}
\tilde{p}_{Q}(u)\log(1+e^{-\epsilon_{Q}(u)})\]
One has to be careful with such an analytical continuation in the
presence of bound-states. Bound-states require pole singularities
of the scattering matrices which usually cross the integration contour
in the analytical continuation and result in extra source terms. See
the Lee-Yang model in the relativistic case \cite{Dorey:1996re,Dorey:1997rb}
for example.

\section{TBA for planar AdS/CFT}

In this section we push forward the TBA program for planar AdS/CFT.
The main difference compared to the previous discussion lies in the
nondiagonal nature of the scattering matrix. There is a way, however,
how we can profit from the previous diagonal results: the nondiagonal
nature of any theory can be encoded into a diagonal theory but with
auxiliary degrees of freedom. These auxiliary excitations do not contribute
to the energy merely modifies the allowed momenta. Let us now follow
the steps of Section 2.

\subsection{Infinite volume characteristics of the model}

The symmetry algebra of the theory has a factorized form: $su(2\vert2)\otimes su(2\vert2)$.
The fundamental particle called magnon transforms in the bifundamental
representation whose S-matrix has the structure
\begin{equation}
\mathbb{S}_{11}(p_{1},p_{2})=S(p_{1},p_{2})\hat{S}_{11}(p_{1},p_{2})
\otimes\hat{S}_{11}(p_{1},p_{2})\label{eq:Smatrix}
\end{equation}
where the matrix part $\hat{S}$ is fixed from its covariance under
one copy of $su(2\vert2)$ up to a scalar factor, which is determined
from unitarity and crossing symmetry. The scattering matrix has simple
poles corresponding to bound-states. There is an infinite tower of
bound-states labeled by a positive integer charge $Q$. They transform
under the tensor product of the atypical totally \emph{symmetric}
representations of the algebra and have dispersion relation\[
E_{Q}(p)=\sqrt{Q^{2}+4g^{2}\sin^{2}\frac{p}{2}}\]

\subsection{Infinite volume characteristics of the mirror model}

As the mirror model is derived from the same Euclidean theory the
fundamental particles' scattering matrix is the analytical continuation
of the scattering matrix (\ref{eq:Smatrix}). We are in a different
analytical domain, however, and here different poles correspond to
bound-states. These bound-states are also labeled by the charge $Q$
but they transform under the atypical totally \emph{antisymmetric}
representations and have dispersion relation:\[
\tilde{E}_{Q}(\tilde{p})=2\mbox{arcsinh}\left(\frac{1}{2g}
\sqrt{\tilde{p}^{2}+Q^{2}}\right)\]

\subsection{Very large volume solution: ABA for the model}

If we put $N$ particles in a finite volume $L$ the momenta of the
particles will be quantized. The multiparticle wave function has to
be periodic in each argument, that is when a particle transported
along the cylinder it scatters on all other particles before arriving
back to its initial position. In a diagonal theory this results in
(\ref{eq:ABA}). In a nondiagonal theory, however, the multiparticle
transfer matrix has to be diagonalized. This can be achieved by introducing
new type of (magnonic) particles with vanishing dispersion relations
and considering the original problem in terms of them as a diagonal
scattering theory. 

Here we focus only on the charge $Q=1$ sector of the theory. We have
momentum carrying particles ($\bullet_{1}$) which scatter on each
other as 
\footnote{The index $1$ in $\bullet_{1}$ refers to the charge of the particle.
This particle is a first member of an infinite series of bound-states
labeled by $\bullet_{Q}$. Similarly we will meet particles of type
$\circ_{N}$ and $\triangleright_{M}$.
}\[
S_{11}^{\bullet\bullet}(p_{1},p_{2})=S(p_{1},p_{2})=
\frac{x_{1}^{+}-x_{2}^{-}}{x_{1}^{-}-x_{2}^{+}}\frac{1-
\frac{1}{x_{1}^{-}x_{2}^{+}}}{1-\frac{1}{x_{1}^{+}x_{2}^{-}}}\,\sigma_{12}^{-2}\]
where $x^{\pm}(p)=\frac{\left(\cot\frac{p}{2}\pm i\right)}{2g}\left(1+
\sqrt{1+4g^{2}\sin^{2}\frac{p}{2}}\right)$
and $\sigma$ represents the dressing phase. These particles are extended
for each $su(2\vert2)$ factor with two types of auxiliary particles
($y,\circ_{1}$), whose parameters are labeled by $y\in\mathbb{R}$
and $w\in\mathbb{R}$. The auxiliary particles have trivial dispersion
relations (their energy and momentum are zero) and scatter with the
fundamental, momentum carrying ones as\[
S_{1y}^{\bullet y}(p,y)=\frac{x^{-}-y}{x^{+}-y}\sqrt{\frac{x^{+}}{x^{-}}}=
S_{y1}^{y\bullet}(y,p)^{-1}\quad;\qquad S_{11}^{\bullet\circ}(p,w)=1\]
 Furthermore, they scatter on each other as\[
S_{11}^{\circ\circ}(w_{1},w_{2})=S_{-2}(w_{1}-w_{2})\quad;\qquad S_{y1}^{y\circ}(y,w)=S_{1}(v(y)-w)
\quad;\qquad S_{yy}^{yy}(y_{1},y_{2})=1\]
 where $v(y)=y+y^{-1}$ and we introduced a useful function 
$S_{n}(v-w)=\frac{v-w+\frac{in}{g}}{v-w-\frac{in}{g}}$.
Any scattering matrix can be extended by unitarity to the opposite
order of their particle types/arguments: $S(i,j)S(j,i)=1$. 

In formulating the ABA equations for the full theory we have to take
into account the two $su(2\vert2)$ factors and that they commute.
The ABA equation for the momentum carrying particles reads as\[
1=e^{ip_{j}L}\prod_{k:k\neq j}^{N_{1}^{\bullet}}S_{11}^{\bullet\bullet}(p_{j},p_{k})
\prod_{\alpha=1,2}\prod_{l=1}^{N_{y^{\alpha}}}S_{1y}^{\bullet y}(p_{j},y_{l}^{\alpha})\]
where $N_{1}^{\bullet}$ is the number of fundamental and $N_{y^{\alpha}}$
the number of $y$ type particles, while the $\alpha=1,2$ index refers
to the two $su(2\vert2)$ factors. Since the two factors commute the
ABA equations for the auxiliary particles with rapidities $y^{1,2}$
and $w^{1,2}$ can be written as\[
\prod_{k:k\neq j}^{N_{1}^{\bullet}}S_{y1}^{y\bullet}(y_{j}^{\alpha},p_{k})
\prod_{l=1}^{N_{1,\alpha}^{\circ}}S_{y1}^{y\circ}(y_{j}^{\alpha},w_{l}^{\alpha})=1=
\prod_{k:k\neq j}^{N_{y^{\alpha}}}S_{1y}^{\circ y}(w_{j}^{\alpha},y_{k}^{\alpha})
\prod_{l:l\neq k}^{N_{1,\alpha}^{\circ}}S_{11}^{\circ\,\circ}(w_{j}^{\alpha},w_{l}^{\alpha})\]
Not all solutions of the ABA equations correspond to single trace
operators as the level matching/zero momentum condition has to be
fulfilled $\sum_{j}p_{j}=0$. The theory contains also bound-states
which can be determined from the singularity structure of the scattering
matrices. Since from the TBA point of view only the bound-state spectrum
of the mirror theory is relevant we will focus only on them.

\subsection{Very large volume solution: ABA for the mirror model}

In the case of the mirror theory the fundamental scattering matrix
is the analytical continuation of the original one $p\to\tilde{p}$.
As a result the ABA will be the analytical continuation, too
\begin{eqnarray}
1 & = & e^{i\tilde{p}_{j}R}\prod_{k:k\neq j}^{N_{1}^{\bullet}}S_{11}^{
\bullet\bullet}(\tilde{p}_{j},\tilde{p}_{k})\prod_{\alpha=1,2}
\prod_{l=1}^{N_{y^{\alpha}}}S_{1y}^{\bullet y}(\tilde{p}_{j},y_{l}^{\alpha})\label{eq:MirrorAdSABA1}\\
-1 & = & \prod_{k:k\neq j}^{N_{1}^{\bullet}}S_{y1}^{y\bullet}(y_{j}^{\alpha},\tilde{p}_{k})
\prod_{l=1}^{N_{1,\alpha}^{\circ}}S_{y1}^{y\circ}(y_{j}^{\alpha},w_{l}^{\alpha})
\label{eq:MirrorAdSABA2}\\ 1 & = & \prod_{k:k\neq j}^{N_{y}^{\alpha}}S_{1y}^{
\circ y}(w_{j}^{\alpha},y_{k}^{\alpha})\prod_{l:l\neq k}^{N_{w}^{\alpha}}S_{11}^{\circ\circ}(w_{j}^{\alpha},w_{l}^{\alpha})\label{eq:MirrorAdSABA3}\end{eqnarray}
There are some differences compared to the original ABA. First the
domain of $\tilde{p}\in\mathbb{R}$ is different compared to $p\in[-\pi,\pi]$
and the total mirror momentum does not need to vanish. Then, as we
are in the mirror theory, the way how $x^{\pm}$ is expressed in terms
of $\tilde{p}$ is also different: $x^{\pm}=\frac{\left(\tilde{p}-i\right)}{2g}
\left(\sqrt{1+\frac{4g^{2}}{1+\tilde{p}^{2}}}\mp1\right)$.
Additionally, in the calculations of the ground state energy the sectors
with antiperiodic fermions are relevant and this is manifested in
a minus sign in the middle equation. The possible bound-states and
their ABA equations are the subject of the next section. Let us note
that usually in the literature instead of (\ref{eq:MirrorAdSABA3})
its inverse is considered as this will lead to positive particle densities
in the thermodynamic limit.

\subsection{Exact groundstate energy: TBA}

In this section we derive TBA integral equations for the groundstate
energy in finite volume $R$. We treat the theory as if it were diagonal
with the scattering matrices specified above. First we analyze whether
this {}``diagonal'' theory has bound-states by analyzing the thermodynamic
behavior of the equations and calculate the scattering matrices of
the bound-states, the so called strings. They are special complex
solutions of the ABA equations and they all contribute to the partition
function which determines the ground state energy. Then we use the
canonical procedure to derive coupled integral equations for the pseudo
energies in a raw form, finally, using identities between the scattering
matrices originating from the symmetry, we rewrite them in a simplified
form and analyze simple excited states.

\subsubsection{String hypothesis for the mirror model}

The string hypothesis is similar to closing the S-matrix bootstrap
program, that is to identify all particles (including bound-states)
of the theory and to determine their scattering matrices. Let us premise
that we will find bound-states of three infinite types $(\bullet_{Q},\triangleright_{M},\circ_{N})$
for $Q,M,N\in\mathbb{N},$ and also of a finite type $y_{\delta}$
particle with $\delta\in\{\pm\}$. They can be arranged in the two
dimensional lattice shown in Figure 1. Let us see how they arise from
the ABA equations. 

In the following we put $R$ and all particle numbers large (keeping
their ratio finite) and analyze the ABA one by one. Let us first note
the reality properties of the equations. Unitarity of the mirror scattering
matrix implies that the $y$ roots come in complex conjugated pairs
$y_{i}=(y_{j}^{-1})^{*}$ or lie on the unit circle $y=(y^{-1})^{*}$,
similarly the roots $w$ come in complex conjugated pairs $w_{i}=w_{j}^{*}$or
are real.

\subsubsection*{$\bullet_{Q}$ particles}

In looking for momentum bound-states we rewrite the scattering matrix
in (\ref{eq:MirrorAdSABA1}) as\[
S_{11}^{\bullet\,\bullet}(\tilde{p}_{1},\tilde{p}_{2})=
\frac{u_{1}-u_{2}+\frac{2i}{g}}{u_{1}-u_{2}-\frac{2i}{g}}\,\Sigma_{11}^{-2}\qquad;
\qquad\Sigma_{11}=\frac{1-\frac{1}{x_{1}^{+}x_{2}^{-}}}{1-\frac{1}{x_{1}^{-}x_{2}^{+}}}\sigma\]
where the rapidity is introduced as $u\pm\frac{i}{g}=x^{\pm}+\frac{1}{x^{\pm}}$.
As $R$ is very large complex values for $u_{1}$ with positive imaginary
part are allowed. In this case the lhs. of (\ref{eq:MirrorAdSABA1})
for $j=1$ diverges so there should be another $u$ say $u_{2}$ that
goes to $u_{1}-\frac{2i}{g}$. If $u_{2}$ still has a positive imaginary
part then by the same argument there should be another $u$ say $u_{3}$
which goes to $u_{2}-\frac{2i}{g}$. Applying this procedure we arrive
at a string of $Q$ roots $u+(Q-1)\frac{i}{g},u+(Q-3)\frac{i}{g},\dots,u-(Q-3)
\frac{i}{g},u-(Q-1)\frac{i}{g}$
or shortly $u_{Q+1-2j}=u+i(Q+1-2j)\frac{i}{g}$ where $j=1,\dots,Q$.
(Clearly the $Q=1$ string is the original particle itself.) The scattering
of the $Q$-string with any other particle of type $(.)$, label $i$
and rapidity $q$ is\[
S_{Qi}^{\bullet\,.}(u,q)=\prod_{j=1}^{Q}S_{1i}^{\bullet\,.}(u_{Q+1-2j},q)=S_{iQ}^{.\,\bullet}(q,u)^{-1}\]
Although naively the scattering matrices seem to depend on the parameters
$x^{\pm}$ and such a way the bound-state scattering matrix depends
on its constituents, this is not the case when we take into account
the contributions of the dressing phase as was shown in \cite{Arutyunov:2009kf}. 

The auxiliary particles exist for both $su(2\vert2$) factors. Here
we focus only on one of them and omit to write out its index.

\subsubsection*{$y_{\delta}$ particles}

Let us analyze (\ref{eq:MirrorAdSABA2}). If we suppose that the number
of momentum carrying particles $N_{1}^{\bullet}$ goes to infinity
then\begin{equation}
\prod_{k:k\neq j}^{N_{1}^{\bullet}}S_{y1}^{y\bullet}(y_{j},\tilde{p}_{k})\to\left\{ \begin{array}{ccc}
0 & \mbox{if} & \vert y_{j}\vert<1\\
\pm1 & \mbox{if} & \vert y_{j}\vert=1\\
\infty & \mbox{if} & \vert y_{j}\vert>1\end{array}\right.\label{eq:ystring}\end{equation}
In the middle case $y$ roots lying on the unit circle are allowed.
As the scattering matrix $S_{y1}^{y\circ}(y,w)$ has a difference
form in the variable $v(y)=y+y^{-1}$ we might use the parameter $v$
instead of $y$. The inverse of the relation, however, is not unique.
Defining $y_{-}(v)=\frac{1}{2}(v-i\sqrt{4-v^{2}})$ with the branch
cuts running from $\pm\infty$ to $\pm2$ we can describe any $y$
with $\Im m(y)<0$ for $v\in[-2,2]$. Clearly $y_{+}(v)=y_{-}(v)^{-1}$
describes the other $\Im m(y)>0$ case and in the scattering matrices
$S_{y1}^{y\bullet}$ which depends on $y$, and not on $v$, we have
to specify which root is taken. As a consequence we have two types
of $y$ particles $y_{\delta}$ with $\delta=\pm$ and the scattering
matrices split as $S_{y1}^{y\bullet}(y,q)\to S_{\delta1}^{y\bullet}(y_{\delta}(v),q)=:
S_{\delta1}^{y\bullet}(v,q)$.

\subsubsection*{$\triangleright_{M}$ particles}

If $\vert y_{1}\vert<1$ in (\ref{eq:ystring}) then the rhs. of (\ref{eq:MirrorAdSABA2})
goes to zero which has to be compensated by a $w_{1}$ root which
goes to $v_{1}-\frac{i}{g}=y_{1}+y_{1}^{-1}-\frac{i}{g}.$ But then
taking the ABA for $w_{1}$ means that the rhs. of (\ref{eq:MirrorAdSABA3})
will diverge which has to be compensated by a root $v_{2}=w_{1}-\frac{i}{g}$.
If the corresponding $y_{2}$ satisfies $\vert y_{2}\vert>1$ then
(\ref{eq:MirrorAdSABA2}) is consistent with (\ref{eq:ystring}) and
reality requires $y_{1}=(y_{2}^{-1})^{*}$, $w_{1}=w_{1}^{*}$. The
three roots $y_{1}\leftrightarrow v_{1}=v+\frac{i}{g}$ and $w_{1}=v$
and $v-\frac{i}{g}=v_{2}\leftrightarrow y_{2}$ form an $M=1$ string
which we denote by $\triangleright_{1}$. In the case when $\vert y_{2}\vert<1$
then we have to repeat the same arguments for $y_{2}$ leading to
$w_{2}$ and $y_{3}$ and so on. Finally we arrive at the notion of
a $\triangleright_{M}$ string. It consists of $2M$ $y$ particles
with $y_{j}=(y_{-j}^{-1})^{*}$ and $M$ $\circ$ particles with synchronized
parameters $w_{M+1-2j}=v+(M+1-2j)\frac{i}{g}$ and $y_{j}\to v_{\mathrm{sign}(j)(M+2-2j)}=v+\mathrm{sign}(j)(M+2-2j)\frac{i}{g}$
for $j=1,\dots,M$. The composite scattering matrix of the $\triangleright_{M}$
particle with all other particles is simply the product of the scatterings
of its each individual constituents\[
S_{Mi}^{\triangleright\,.}(v,q)=\prod_{j=1}^{M+1}S_{-i}^{y\,.}(v_{M+2-2j},q)\prod_{j}^{M}S_{1i}^{\circ\,.}(w_{M+1-2j},q)\prod_{j=1}^{M-1}S_{+i}^{y\,.}(v_{M-2j},q)=S_{iM}^{.\,\triangleright}(i,v)^{-1}\]

\subsubsection*{$\circ_{N}$ particles}

Suppose we have a large number of $y$ particles and that $w_{1}$
has a positive imaginary part. Then the first factor of the rhs. of
(\ref{eq:MirrorAdSABA3}) will go to zero which has to be compensated
by a root $w_{2}=w_{1}-\frac{2i}{g}$. If $\Im m(w_{2})<0$ then we
obtain a $\circ_{2}$ string. In the opposite case we repeat to previous
argumentation leading to an $N$ string $w_{N+1-2j}=w+(N+1-2j)\frac{i}{g}$.
Clearly a single $w$ is just a $\circ_{1}$ string. The scattering
of the $N$ string with any other particle is\[
S_{Ni}^{\circ\,.}(w,i)=\prod_{j=1}^{N}S_{wi}^{\circ\,.}(w_{N+1-2j},i)\]

\subsubsection*{Scattering matrices }

Summarizing, the mirror AdS theory in the thermodynamic limit could
be replaced by a diagonal theory having constituents of infinite type
$(\bullet,\triangleright,\circ)$ and index $Q,M,N$ for $Q,M,N\in\mathbb{N},$
and also of finite type $y$ particles with $\delta\in\{\pm\}$. See
also Figure 1. 

\begin{table}[h]
\begin{centering}
\begin{tabular}{|c||c|c|c|c|}
\hline 
 & $\bullet_{Q'}$ & $\triangleright_{M'}$ & $\circ_{N'}$ & $y_{\delta'}$\tabularnewline
\hline
\hline 
$\bullet_{Q}$ & $S_{QQ'}^{\bullet\,\bullet}$ & $S_{QM'}^{\bullet\,\triangleright}$ & $1$ &
 $S_{Q\delta'}^{\bullet\, y}$\tabularnewline
\hline 
$\triangleright_{M}$ & $S_{MQ'}^{\triangleright\,\bullet}$ & $S_{MM'}^{\triangleright\,
\triangleright}$  & $1$ & $S_{M\delta'}^{\triangleright\, y}$\tabularnewline
\hline 
$\circ_{N}$ & $1$ & $1$ & $S_{NN'}^{\circ\,\circ}$ & $S_{N\delta'}^{\circ\, y}$\tabularnewline
\hline 
$y_{\delta}$ & $S_{\delta Q'}^{y\,\bullet}$ & $S_{\delta M'}^{y\,\triangleright}$ & $S_{\delta N'}^{y\,\circ}$ & $1$\tabularnewline
\hline
\end{tabular}
\par\end{centering}

\caption{Scattering matrices of the various particles}
\label{Flo:SmatrixTable}
\end{table}

For the readers convenience we summarize the scattering matrices in
Table \ref{Flo:SmatrixTable}. The scattering matrices are unitary
$S_{ij}S_{ji}=1$ and their explicit forms are
\begin{eqnarray*}
S_{QQ'}^{\bullet\,\bullet}(u,u') & = & S_{QQ'}(u-u')\Sigma_{QQ'}(u,u')^{-2}\\
S_{QQ'}(u-u') & = & S_{Q+Q'}(u-u')S_{Q'-Q}(u-u')\prod_{j=1}^{Q-1}S_{Q'-Q+2j}(u-u')^{2}\\
\Sigma_{QQ'}(u,u') & = & \prod_{j=1}^{Q}\prod_{k=1}^{Q'}\sigma(u_{Q+1-2j},u_{Q'+1-2k})
\frac{1-\frac{1}{x(u_{Q-2j})x(u_{Q'+2-2k})}}{1-\frac{1}{x(u_{Q+2-2j})x(u_{Q'-2k})}}\end{eqnarray*}
where $u_{j}=u+j\frac{i}{g}$ and we reparametrized the momentum carrying
particles in terms of the rapidity via the function $x(u)=\frac{1}{2}(u-i\sqrt{4-u^{2}})$.
Recall also that $S_{n}(u-w)=\frac{u_{n}-w}{u_{-n}-w}$. The other
matrix elements are\begin{eqnarray*}
S_{QM}^{\bullet\,\triangleright}(u,v) & = & \frac{x(u_{-Q})-x(v_{M})}{x(u_{Q})-x(v_{M})}\,
\frac{x(u_{-Q})-x(v_{-M})}{x(u_{Q})-x(v_{-M})}\,\frac{x(u_{Q})}{x(u_{-Q})}\prod_{j=1}^{M-1}
S_{M-Q-2j}(u,v)\\ S_{Q\delta}^{\bullet\, y}(u,v) & = & \frac{x(u_{-Q})-x(v)^{\delta}}{x(u_{Q})-x(v)^{\delta}}\sqrt{\frac{x(u_{Q})}{x(u_{-Q})}}\\
S_{MM'}^{\triangleright\,\triangleright}(u,u) & = & S_{MM'}(u-u')=\; S_{MM'}^{\circ\,\circ}(u,u')^{-1}\\
S_{M\delta}^{\triangleright\, y}(u,v) & = & S_{M}(u-v)=\; S_{M\delta}^{\circ\, y}(u,v)\end{eqnarray*}
The ABA equations then have a generic form
\begin{eqnarray*}
(-1)^F =e^{i\tilde{p}_{.}(q_{j})R}\prod_{k}S_{jQ_{k}}^{.\,\bullet}(q_{j},u_{Q_{k}})
\prod_{\alpha=1,2}\prod_{l}S_{j\delta_{l}}^{.\, y}(q_{j},v_{l}^{\alpha})
\prod_{m}S_{jM_{m}}^{.\,\triangleright}(q_{j},v_{M_{m}}^{\alpha})
\prod_{n}S_{jN_{n}}^{.\,\circ}(q_{j},w_{N_{n}})
\end{eqnarray*}
where $.$ can be any type of $\bullet,\triangleright,\circ,y$ but
only the $\bullet$ particles have nonvanishing energy $\tilde{E}_{Q}$
and momentum $\tilde{p}_{Q}(u)=g(x(u_{-Q})-x(u_{Q}))+iQ$. The parameter
$F$ denotes the fermion number. We also indicated the contributions
of the two $su(2\vert2)$ factors. The energy of such a multiparticle
state having $N_{Q_{k}}^{\bullet}$ of $Q_{k}$ particles is\[
\tilde{E}(\tilde{p}_{1},\dots,\tilde{p}_{k})=\sum_{k}\tilde{E}_{Q_{k}}(\tilde{p}_{k})\]
Let us note that the ABA equations for the auxiliary particles can
be inverted without changing their physical meaning. Taking the inverse
of (\ref{eq:MirrorAdSABA3}) is equivalent to redefining simultaneously
the scattering matrices $S_{1\delta}^{\circ y}\to(S_{1\delta}^{\circ y})^{-1}$
and $S_{11}^{\circ\circ}\to(S_{11}^{\circ\circ})^{-1}$. Actually
these are the equations used in the literature as they give positive
particle densities in the thermodynamic limit.

\subsubsection{Raw TBA equations }

Suppose now that we would like to describe the groundstate energy
in the AdS system in volume $L$. In doing so we follow the steps
presented in Section 2 to evaluate the partition function for large
mirror sizes. We introduce densities of particles (strings) $\rho_{Q}^{\bullet}(u),$
$\rho_{M}^{\triangleright}(u),$ $\rho_{N}^{\circ}(u)$ for $u\in\mathbb{R}$
and $\rho_{\delta}^{y}(u)$ for $u\in[-2,2]$ and the analogous densities
of holes $\rho\to\bar{\rho}.$ They are restricted via the logarithm
of the ABA which contains the logarithmic derivatives of the scattering
matrices\[
K_{jj'}^{\,.\,.}(u,u')=-i\partial_{u}\log S_{jj'}^{\,.\,.}(u,u')\]
Clearly $K_{jj'}^{\,.\,.}(u,u')\neq-K_{j'j}^{\,.\,.}(u',u)$ as the
scattering matrices are not of the difference type. (Keeping in mind
how we obtained the string solutions the densities are naturally ordered
$\rho_{Q}^{\bullet}\gg\rho^{y}\gg\rho_{N}^{\circ},\rho_{M}^{\triangleright}$.)
Then we introduce the entropy factors for the densities, $i\pi$ chemical
potential for fermions and calculate the saddle point of the functional
integral. This results in integral equations for the pseudo energies
$\epsilon_{Q}^{\bullet},\epsilon_{M}^{\triangleright},\epsilon_{N}^{\circ},\epsilon_{\delta}^{y}$
as follows\[
\epsilon_{Q}^{\bullet}=L\tilde{E}_{Q}-\log(1+e^{-\epsilon_{Q'}^{\bullet}})\star K_{Q'Q}^{\bullet\,\bullet}-\log(1+e^{-\epsilon_{M}^{\triangleright}})\star K_{MQ}^{\triangleright\,\bullet}-\log(1+e^{-\epsilon_{\delta}^{y}})\star K_{\delta Q}^{y\,\bullet}\]
where in the contributions of the $\triangleright_{M}$ and $y_{\delta}$
we have to sum for the contributions of the two $su(2\vert2$) factors
(which we omitted to write out). The remaining equations are valid
separately for the two $su(2\vert2)$ factors separately:
\begin{eqnarray*}
\epsilon_{M}^{\triangleright} & = & -\log(1+e^{-\epsilon_{Q}^{\bullet}})\star K_{QM}^{\bullet\,\triangleright}-\log(1+e^{-\epsilon_{M'}^{\triangleright}})
\star K_{M'M}^{\triangleright\,\triangleright}-\log(1+e^{-\epsilon_{\delta}^{y}})\star K_{\delta M}^{y\,\triangleright}\\ \epsilon_{N}^{\circ} & = & \log(1+e^{-\epsilon_{N'}^{\circ}})
\star K_{N'N}^{\circ\circ}+\log(1+e^{-\epsilon_{\delta}^{y}})\star K_{\delta N}^{y\,\circ}\\
\epsilon_{\delta}^{y} & = & -\log(1+e^{-\epsilon_{Q}^{\bullet}})\star K_{Q\delta}^{\bullet\, y}-
\log(1+e^{-\epsilon_{M}^{\triangleright}})\star K_{M\delta}^{\triangleright\, y}-\log(1+e^{-
\epsilon_{N}^{\circ}})\star K_{N\delta}^{\circ\, y}+i\pi\end{eqnarray*}
Once these equations are solved the groundstate energy can be obtained
as\[
E_{0}(L)=-\sum_{Q=1}^{\infty}\int\frac{du}{2\pi}\partial_{u}\tilde{p}_{Q}\log(1+e^{-\epsilon_{Q}^{\bullet}})\]

Finally we note that we replaced the magnonic ABA for the particle
type $\circ_{N}$ with its inverse and made the corresponding change
in the scattering matrices to ensure the positivity of the magnonic
densities $\rho_{N}^{\circ}$. It effectively changed the sign of
the related kernels.

\subsubsection{Simplified TBA equations and Y-system }

In this subsection using identities among the TBA kernels we bring
the equations in to a universal local form. This means that pseudo
energies can be drawn in a two dimensional lattice, such that only
neighboring sites couple to each other with the following universal
kernel\begin{equation}
s\, I_{MN}=\delta_{MN}-(K+1)_{MN}^{-1}\quad;\qquad s(u)=\frac{g}{4\cosh\frac{g\pi u}{2}}\label{eq:Ksop}\end{equation}
where $I_{MN}=\delta_{M+1,N}+\delta_{M-1,N}$ and $(K+1)_{MN}^{-1}\star(K_{NL}+\delta_{NL})=\delta_{ML}$.
To simplify the notation let us introduce the following $Y$ functions\[
Y_{Q}^{\bullet}=e^{-\epsilon_{Q}^{\bullet}}\quad;\quad Y_{M}^{\triangleright}=e^{-\epsilon_{M}^{\triangleright}}\quad;\quad Y_{N}^{\circ}=e^{\epsilon_{N}^{\circ}}\quad;\quad Y_{\delta}^{y}=e^{\delta\epsilon_{\delta}^{y}}\]
Clearly we have two copies for $Y_{M}^{\triangleright,\alpha},Y_{N}^{\circ,\alpha},Y_{\delta}^{y,\alpha}$.
(To conform with the literature we inverted the ABA equations for
$\triangleright_{M}$ and $y_{-}$). Acting with the operator (\ref{eq:Ksop})
on these inverted TBA equations and using kernel identities like $(K+1)_{MN}^{-1}\star K_{N}=s\,\delta_{M,1}$
we arrive at their simplified, universal form\begin{eqnarray*}
\log Y_{M}^{\triangleright} & = & \log(1+Y_{M+1}^{\bullet})\star s-I_{MM'}\log(1+\frac{1}{Y_{M'}^{\triangleright}})\star s+\delta_{M,1}\log\frac{1+Y_{+}^{y}}{1+\frac{1}{Y_{-}^{y}}}\hat{\star}s\\
\log Y_{N}^{\circ} & = & I_{NN'}\log(1+Y_{N'}^{\circ})\star s+\delta_{N,1}\log\frac{1+Y_{-}^{y}}{1+\frac{1}{Y_{+}^{y}}}\hat{\star}s\end{eqnarray*}
where in the convolution $\hat{\star}$ we integrate over the interval
$[-2,2]$ only. The other equations do not behave so nicely. \begin{eqnarray*}
\log Y_{Q}^{\bullet} & = & -I_{QQ'}\log(1+\frac{1}{Y_{Q'}^{\bullet}})\star s+\log(1+Y_{Q-1}^{\triangleright,1})\star s+\log(1+Y_{Q-1}^{\triangleright,2})\star s\;;\quad Q>1\\
\log Y_{1}^{\bullet} & = & -\log(1+\frac{1}{Y_{2}^{\bullet}})\star s+(\log(1+Y_{-}^{y,1})(1+Y_{-}^{y,2}))\star s-\check{\Delta}\star s\end{eqnarray*}
where $\check{\Delta}$ vanishes on the interval $[-2,2]$ whose explicit
form can be found in \cite{Arutyunov:2009ux}. The equation for the
$y$ particles are simpler in the original form\[
\delta\log Y_{\delta}^{y}=-\log(1+Y_{Q}^{\bullet})\star K_{Q\delta}^{\bullet y}+\log\frac{1+Y_{M}^{\triangleright}}{1+\frac{1}{Y_{M}^{\circ}}}\star K_{M}+i\pi\]
These equations for $Y_{\delta}^{y}$ are not in a local form. However,
acting with the inverse of $s$ they can be brought into such form.
The operator $s^{-1}$ acts as $(f\star s^{-1})(u)=f(u+\frac{i}{g}-i0)+f(u-\frac{i}{g}+i0)$
and involves the analytical continuation of the functions. It has
a large null space, thus when acting on the equation information is
lost: \[
\log Y_{-}^{y}\star s^{-1}=\log(1+Y_{1}^{\bullet})+\log(1+Y_{1}^{\circ})-\log(1+\frac{1}{Y_{1}^{\triangleright}})\]
The advantage of defining $s^{-1}$ in the above manner is that it
uses the analytically continued values of the $Y$ functions on the
rapidity torus only. If we continue them across the cuts by using
$(f\star s^{-1})(u)=f(u+\frac{i}{g}-i0)+f(u-\frac{i}{g}-i0)=f^{+}(u)+f^{-}(u)$
then the term $\check{\Delta}$ disappears, but the $Y$ functions
have to be extended to an infinite genus Riemann surface.  
On this surface the Y-system has the universal
form\begin{equation}
Y_{N,M}^{+}Y_{N,M}^{-}=\frac{(1+Y_{N,M+1})(1+Y_{N,M-1})}{(1+Y_{N-1,M}^{-1})(1+Y_{N+1,M}^{-1})}\label{eq:Ysystem}\end{equation}
 where the $N,M$ indices live on a two dimensional integral lattice.
In our situation the identification can be drawn on Figure 1, which
explicitly reads as $Y_{Q}^{\bullet}=Y_{Q,0}$, $Y_{M}^{\triangleright,\alpha}=Y_{M+1,\nu_{\alpha}}$,
$Y_{N}^{\circ,\alpha}=Y_{1,\nu_{\alpha}(N+1)}$, $Y_{-}^{y,\alpha}=Y_{1,\nu_{\alpha}}$
and $Y_{+}^{y,\alpha}=Y_{2,\nu_{\alpha}2}$ where $\nu_{1}=1$ and
$\nu_{2}=-1$. 

\begin{figure}
\begin{centering}
\includegraphics[width=6cm]{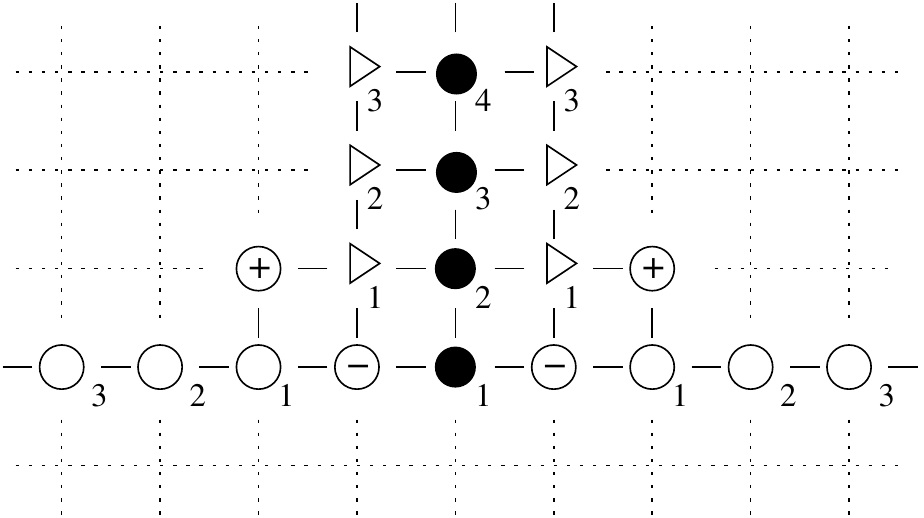}
\par\end{centering}

\caption{$Y$- system for planar AdS/CFT. $Y_{-}$ is denoted by $\ominus$
while $Y_{+}$ by $\oplus$. }

\end{figure}

\subsubsection{Excited states by analytical continuation}

Here we focus on the TBA equations for excited states in the $sl_{2}$
sector for small coupling. This sector contains particles of type
$\bullet_{1}$ only and have ABA:\[
1=e^{ip_{k}L}\prod_{j:j\neq k}S_{11}^{\bullet\,\bullet}(p_{k},p_{l})\]
These equations are asymptotic only and the exact system of TBA equations
is required to describe the energy of the multiparticle state exactly.
As the vacuum is a BPS state it has vanishing energy and its analytical
continuation cannot describe excited states. Alternatively we choose
an integration contour, such that when it is taken back to the real
axis the residue of a singularity of the form $1+e^{-\epsilon_{1}^{\bullet}(p_{k})}=0$
is picked up resulting in additional source terms in the raw equations
as:\[
\epsilon_{Q}^{\bullet}\to\sum_{j}\log S_{1Q}^{\bullet\,\bullet}(p_{j},u)\quad;\qquad\epsilon_{M}^{\triangleright}\to\sum_{j}\log S_{1M}^{\bullet\,\triangleright}(p_{j},u)\quad;\qquad\epsilon_{\delta}^{y}\to\sum_{j}\log S_{1\delta}^{\bullet\, y}(p_{j},u)\]
Once the new system of TBA equations are solved the pseudo energies
$\epsilon_{Q}^{\bullet}$ have to be plugged into the energy formula:\[
E(L)=\sum_{k}E_{1}(p_{k})-\sum_{Q=1}^{\infty}\int\frac{du}{2\pi}\partial_{u}\tilde{p}_{Q}\log(1+e^{-\epsilon_{Q}^{\bullet}})\]
to obtain the energy of the multiparticle system. 

We can rewrite the TBA equations in terms of the $Y$ functions into
their simplified form. They satisfy the same $Y$-system relations
(\ref{eq:Ysystem}) but with a different asymptotical behavior. There
is a systematical asymptotical expansion of the $Y$-system, which
reproduces both the ABA and the leading L\"uscher correction of these
multiparticle states. This is valid for weak coupling $g\to0$ (or
large sizes) and it is very nontrivial to follow the analytical behavior
of the $Y$ functions as one increases the coupling. The ABA solution
itself suggests, that additional $1+Y=0$ singularities could appear
and then the TBA equations have to be modified by additional source
terms. These source terms ensure the analytical behavior of the energy
around these singular points.

\section{Guide to the literature}

Here we list the representative papers where the various parts of
the TBA program were developed.

The idea that the TBA program can be applied in the planar AdS/CFT
setting was presented in \cite{Ambjorn:2005wa}. The infinite volume
scattering description of theory can be found in chapters \cite{chapSMat,chapSProp}.
The ABA equations for the planar AdS/CFT model was conjectured in
\cite{Beisert:2005fw} (and thoroughly discussed in chapters \cite{chapLR,chapABA}),
while the analogous ABA for the mirror model was described in \cite{Arutyunov:2007tc}.
As the color structure ($su(2\vert2))$ of the scattering matrix is
the same as that of the Hubbard model, the Hubbard TBA solution can
be adopted \cite{Hubbardbook}. This results in the string hypothesis
which was formulated explicitly in \cite{Arutyunov:2009zu}.
The standard procedure leads to raw TBA equations, which were developed
in \cite{Arutyunov:2009ur,Bombardelli:2009ns,Gromov:2009bc}. The
simplified form of the TBA equations was presented in \cite{Arutyunov:2009ux}
and the Y-system relations, presented previously in \cite{Gromov:2009tv},
were derived in \cite{Arutyunov:2009ur,Bombardelli:2009ns,Gromov:2009bc}.
In doing this the analytical properties of the dressing phase \cite{Arutyunov:2009kf,Gromov:2009bc,Volin:2009uv}
had to be investigated. In the AdS/CFT context the volume of the integrable
system has to be an integer, which can be seen also on the groundstate
TBA \cite{Frolov:2009in}. 

Although we obtained the Y-system from the ground-state TBA equations,
in principle, it follows from the hidden $PSU(2,2\vert4)$ symmetry
of the model. An independent alternative approach based on this symmetry
is the subject of the next Chapter in this volume \cite{chapTrans}.

The Y-system plays a crucial role in describing excited states. As
it is related to the symmetry of the model \cite{chapTrans,Gromov:2010vb,Volin:2010xz}
it is the same for each state. What makes the difference is the asymptotical
and analytical behavior of the Y-functions. The analytical properties
of the  Y-functions was thoroughly analyzed in \cite{Arutyunov:2009ur,Arutyunov:2009ax,Frolov:2009in,Cavaglia:2010nm}.
Based on the solution of the Y-system of the $O(4)$ model \cite{Gromov:2008gj}
the authors of \cite{Gromov:2009tv} identified the large volume solution
in terms of the transfer matrices of the ABA \cite{chapABA}. This
helps to derive excited states TBA equations for the $sl_{2}$ sector,
which was done in \cite{Gromov:2009bc,Arutyunov:2009ax}. The excited
state TBA equations provide an exact description of the given state
and they were used in the Konishi case, \cite{Gromov:2009zb,Frolov:2010wt},
to analyze numerically the behavior of the energy for large coupling.
The results are summarized in Figure 2%
\footnote{We thank the authors of \cite{Gromov:2010km} for borrowing their
figure.%
}, see also \cite{Gromov:2010km}. It was further shown in \cite{Arutyunov:2009ax}
how to modify these excited state TBA equations if a $1+Y=0$ singularity
appears in the analytical continuation in $g$. 

\begin{figure}
\begin{centering}
\includegraphics[width=10cm]{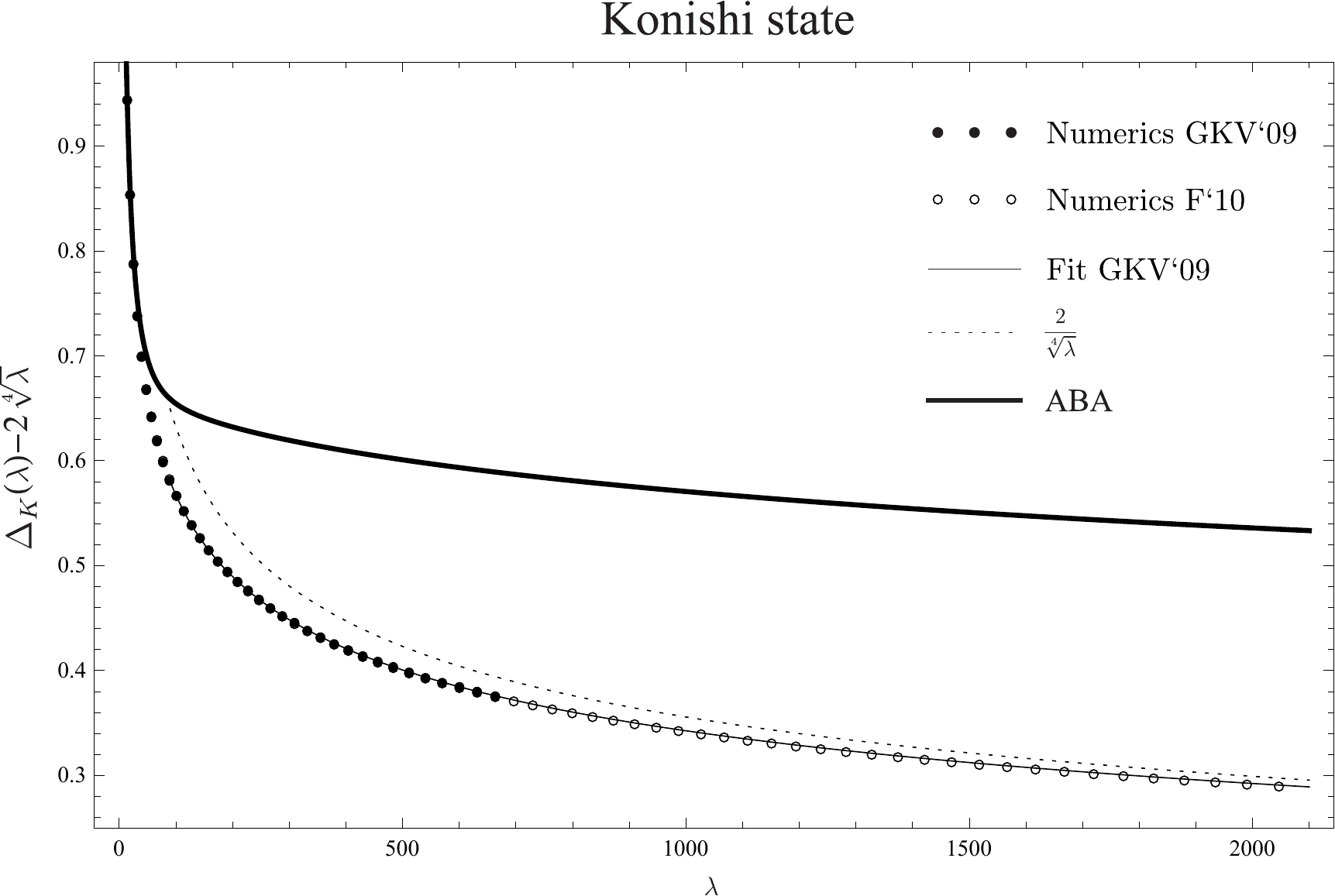}
\par\end{centering}

\caption{Numerical solution of the excited TBA equations for the Konishi state
\cite{Gromov:2009zb,Frolov:2010wt}.}

\end{figure}

The weak coupling limit of the Y-system equations can be compared
to the ABA \cite{chapABA} and L\"uscher type correction \cite{chapLuescher}.
The leading order behavior is built in the asymptotic solution \cite{Gromov:2009tv}
of the Y-function, but the next to leading one provides a stringent
test of the excited states TBA equations, which was performed numerically
for the Konishi operator in \cite{Arutyunov:2010gb} and analytically
at next to leading order in \cite{Balog:2010xa}. Later this analytical
calculation was extended to describe the next to leading order L\"uscher
correction of generic twist two states \cite{Lukowski:2009ce} in
\cite{Balog:2010vf}. 

The strong coupling limit of the $Y$-system for a finite density
of string particles was analyzed in \cite{Gromov:2009tq}, where a
complete agreement with the one loop string energies including all
exponential finite size corrections has been found. The functional
$Y$-system equations were encoded into simpler $Q$ functions in
\cite{Hegedus:2009ky,Gromov:2010vb,Gromov:2010km}. 

\begin{table}[h]
\begin{centering}
\begin{tabular}{|c||c|c|c|c|}
\hline 
$Y_{a,s}$ & This review & AF & BFT & GKKV\tabularnewline
\hline
\hline 
$Y_{Q,0}(u)$ & $Y_{Q}^{\bullet}(u)$ & $Y_{Q}(u)$ & $Y_{Q}(u)$ & $Y_{\bullet_{Q}}(u')$\tabularnewline
\hline 
$Y_{M+1,1}(u)$ & $Y_{M}^{\triangleright}(u)$ & $Y_{M\vert vw}^{-1}(u)$ & $Y_{v,M}(u)$ & $Y_{\triangle_{M+1}}(u')$ \tabularnewline
\hline 
$Y_{1,N+1}(u)$ & $Y_{N}^{\circ}(u)$ & $Y_{N\vert v}(u)$ & $Y_{w,N}(u)$ & $Y_{\circ_{N+1}}(u')$\tabularnewline
\hline 
$Y_{2,2}(u)$ & $Y_{+}^{y}(u)$ & $-Y_{+}(u),$ & $Y_{y}(u)$ & $Y_{\oplus}(u')$\tabularnewline
\hline 
$Y_{1,1}(u)$ & $Y_{-}^{y}(u)$ & $-Y_{-}^{-1}(u),$ & $Y_{y^{*}}(u)$ & $Y_{\otimes}(u')$\tabularnewline
\hline
\end{tabular}
\par\end{centering}

\caption{Relating the Y-functions to those in the literature, where $u'=gu.$ }
\label{Flo:Yfunctions}
\end{table}

Let us mention, how our TBA equations are related to those in the
literature. We summarized the relation between the various conventions
for half of the Y-system in Table 2 as the other half is trivially
related, see also \cite{Cavaglia:2010nm}. Under this replacement
our simplified equations are equivalent to AF \cite{Arutyunov:2009ux},
while the raw equations to BFT \cite{Bombardelli:2009ns}, except for the chemical potentials
of \cite{Bombardelli:2009ns}.
 In comparing to GKKV \cite{Gromov:2009bc} the indentification is not enough. Comparing
our kernel $K_{QQ'}^{\bullet\,\bullet}$ to the one $K_{\bullet_{Q}\bullet_{Q'}}$
in \cite{Gromov:2009bc} we observe a slight difference. This is irrelevant,
however, for excited states satisfying the level matching/zero momentum
condition %
\footnote{We thank the authors of \cite{Gromov:2009bc} for pointing out this. %
}. 

The $AdS_{5}/CFT_{4}$ correspondence has a brother theory, the $AdS_{4}/CFT_{3}$
duality \cite{chapN6}, where the TBA program has been developed
in an analogous way. The ABA together with the string hypothesis of
the mirror theory lead to ground state TBA equations and Y-system
relations in \cite{Bombardelli:2009xz,Gromov:2009at} and extend the
previously conjectured Y-system proposal of \cite{Gromov:2009tv}.
This program is further elaborated in \cite{Gromov:2009at} by additionally
determining excited states TBA equations and comparing them to the
asymptotic solution of the Y functions \cite{Gromov:2009tv} and to
the quasi classical string spectrum. 

Finally, let us list some open problems. 

There are two disagreeing string theory calculations (\cite{Arutyunov:2005hd}
and \cite{Roiban:2009aa}) for the anomalous dimension of the Konishi
state. Additionally, the numerical solution of the TBA equations for
large couplings \cite{Gromov:2009zb,Frolov:2010wt} provides a third
result, and calls for improvements both the string theory and the
TBA sides. On the string theory side it could be a pure spinor calculation,
while on the TBA side one should analyze the analytical behavior of
the Y-system and check whether, with increasing $g$, a singularity
of type $1+Y=0$ indeed appears, as the asymptotic solution suggests
\cite{Arutyunov:2009ax}. In principle the effect of such singularities
is to make the coupling dependence of the energies analytical, but
it has to be established concretely. 

The anomalous dimensions of twist operators in the planar limit can
be described by integral equations derived directly from the ABA \cite{chapTwist}.
It would be nice to see, how the exact excited TBA equations reduce
to these equations in the large spin limit. 

The analytical comparision of the excited state TBA equations to the
next to leading order L\"uscher corrections \cite{Balog:2010xa,Balog:2010vf}
tested explicitly only the $\triangleright$ part of the Y-system.
A next to next to leading order analysis could test the $\circ$ part
as well. 

The excited states TBA equations are coupled nonlinear integral equations
for infinite unknowns. An ideal system of equations should contain
finite unknowns only, and could be developed in analogy to \cite{Gromov:2008gj,Kazakov:2010kf}
by exploiting the result of \cite{Hegedus:2009ky,Gromov:2010vb,Gromov:2010km}.

\subsection*{Acknowledgments}

We thank G. Arutyunov, N. Beisert, D. Fioravanti, S. Frolov, N. Gromov,
A. Hegedus, V. Kazakov, R. Suzuki and R. Tateo for useful comments
on the manuscript. The work was supported by a Bolyai Scholarship,
and by OTKA K81461.

\subsection*{Note added in proof}

After this review chapter was finished three string theory calculations
based on different methods determined the strong coupling expansion
of the anomalous dimension of the Konishi operator \cite{Gromov:2011de,Roiban:2011fe,Vallilo:2011fj}.
All agreed with each other and with the strong coupling expansion
of the TBA equation \cite{Gromov:2009zb,Frolov:2010wt}. This gives
a strong support not only for the correctness of the TBA equations
but also for the integrability approach to planar AdS/CFT.

%%%%%%%%%%%%%%%%%%%%%%%%%%%%%%%%%%%%%%%%
\phantomsection
\addcontentsline{toc}{section}{\refname}
%bibliography generated by nb.bst v1.01 (C) 2003-2010 Niklas Beisert

\end{document}